\documentclass[sigconf, preprint]{acmart}
\usepackage{makecell}
\usepackage{bbm}
\usepackage{bm}
\usepackage{color}
\usepackage{arydshln}
\usepackage{graphicx} 
\usepackage{subfigure} %
\usepackage{array}
\usepackage{amsmath}  
\usepackage{multirow} 
\usepackage{booktabs} 
\usepackage{adjustbox}
\AtBeginDocument{%
  }

\renewcommand{\footnotetextcopyrightpermission}[1]{} 
\begin{document}

\title{Beyond Existing Retrievals: Cross-Scenario Incremental Sample Learning Framework}

\author{Tao Wang}
\email{wt439443@taobao.com}
\orcid{0000-0001-7354-8606}
\affiliation{%
  \institution{Taobao \& Tmall Group of Alibaba}
  \city{Hangzhou}
  \country{China}
}

\author{Xun Luo}
\authornote{Work done during internship at Alibaba.}
\email{luoxun.luo@taobao.com}
\affiliation{%
  \institution{Taobao \& Tmall Group of Alibaba}
  \city{Hangzhou}
  \country{China}}
  
\author{Jinlong Guo}
\authornote{Corresponding author.}
\email{guojinlong.gjl@taobao.com}
\affiliation{%
  \institution{Taobao \& Tmall Group of Alibaba}
  \city{Hangzhou}
  \country{China}}

\author{Yuliang Yan}
\email{yuliang.yyl@taobao.com}
\affiliation{%
  \institution{Taobao \& Tmall Group of Alibaba}
  \city{Hangzhou}
  \country{China}}

\author{Jian Wu}
\email{joshuawu.wujian@alibaba-inc.com}
\affiliation{%
  \institution{Taobao \& Tmall Group of Alibaba}
  \city{Beijing}
  \country{China}}  

\author{Yuning Jiang}
\email{mengzhu.jyn@alibaba-inc.com}
\affiliation{%
  \institution{Taobao \& Tmall Group of Alibaba}
  \city{Beijing}
  \country{China}}

\author{Bo Zheng}
\email{bozheng@alibaba-inc.com}
\affiliation{%
  \institution{Taobao \& Tmall Group of Alibaba}
  \city{Beijing}
  \country{China}}

\renewcommand{\shortauthors}{Tao Wang et al.}
\begin{abstract}
  The parallelized multi-retrieval architecture has been widely adopted in large-scale recommender systems for its computational efficiency and comprehensive coverage of user interests. Many retrieval methods typically integrate additional cross-scenario samples to enhance the overall performance ceiling. However, those model designs neglect the fact that a part of the cross-scenario samples have already been retrieved by existing models within a system, leading to diminishing marginal utility in delivering incremental performance gains.
  In this paper, we propose a novel retrieval framework \textbf{IncRec}, specifically for cross-scenario incremental sample learning. The innovations of IncRec can be highlighted as two aspects. Firstly, we construct extreme cross-scenario incremental samples that are not retrieved by any existing model. And we design an incremental sample learning framework which focuses on capturing incremental representation to improve the overall retrieval performance. Secondly, we introduce a consistency-aware alignment module to further make the model prefer incremental samples with high exposure probability.
  Extensive offline and online A/B tests validate the superiority of our framework over state-of-the-art retrieval methods. In particular, we deploy IncRec in the Taobao homepage recommendation, achieving a 1\% increase in online transaction count, demonstrating its practical applicability.
\end{abstract}




\maketitle

\section{Introduction}
Modern recommender systems have emerged as an indispensable component of the digital ecosystem infrastructure, effectively addressing the "information overload" dilemma and improving user engagement by connecting users with personalized content (e.g., Taobao, Amazon, TikTok, Facebook) ~\cite{rs1,rs2,rs3,rs4,rs5}. Industrial recommender systems typically adopt a cascaded two-stage architecture: the retrieval stage (formally termed candidate generation) and the ranking stage ~\cite{retrieval1,rank1}. This paper focuses on the retrieval stage, which aims to fetch thousands of items from a large-scale candidate pool and establishes the theoretical performance ceiling. 

To achieve computational efficiency and better capture diverse user interests,  large-scale retrieval systems commonly adopt the multi-retrieval architecture, typically including collaborative filtering methods ~\cite{cf1, cf2,cf3}, graph-based methods ~\cite{gnn1,gnn2,gnn3}, and embedding-based dual-tower methods ~\cite{twotower1, twotower2, towtower3}. Most of them focus on intra-scenario prediction optimization. To further improve the performance ceiling, many methods introduce additional cross-scenario samples ~\cite{cross_scenario1,cross_scenario2,cross_scenario3,lyu2023entirespacelearningframework,OTT,BOMGraph}. For example, ESLM ~\cite{lyu2023entirespacelearningframework} utilizes entire space samples to alleviate the data sparsity problem. M5~\cite{OTT} introduces scenario indicators as features and employs split mixture-of-experts to generate embeddings. BOMGraph ~\cite{BOMGraph} captures heterogeneous information flow across scenarios by inter-scenario and intra-scenario metapaths. 

However, these cross-scenario methods still suffer from a critical limitation. They mix cross-scenario samples in a simple manner and focus on designing different model structures to achieve the transfer of information from other scenarios to their primary service scenario. They neglect the fact that a part of the cross-scenario samples have already been retrieved by existing models within a multi-retrieval system because of some common interests across different scenarios, resulting in high redundancy and limited performance. More importantly, they exhibit diminishing marginal utility in producing incremental performance for entire retrieval. 

To solve this limitation, we propose a novel retrieval framework \textbf{IncRec}, focusing on learning cross-scenario incremental samples that are not retrieved by any existing model. From the perspective of sample, we construct cross-scenario samples and separate them into two groups depending on whether they are retrieved by existing models. From the perspective of model, we adopt a basic dual-tower model to learn effective representations from both groups, and introduce another incremental learning tower to extract incremental user representations via explicit contrast between different groups. In addition, we introduce a consistency-aware alignment tower to align with the preference of subsequent ranking via consistency objective. Then, alignment capability can be transferred to incremental learning tower by implicit adaptive weight.
In summary, the principal contributions of this paper are as follows:
\begin{itemize}
\item {We propose a novel retrieval framework to capture incremental representations from cross-scenario incremental samples, which can significantly enhance the overall retrieval performance.}
\item {We propose an alignment module which make the model prefer incremental samples that are more consistent with the preference of subsequent ranking, further enhancing the exposure probability.}
\item {Extensive offline evaluations coupled with online A/B tests deployed in Taobao demonstrate the significant performance and practical applicability of the proposed method.}
\end{itemize}

\section{Methodology}
In this section, we introduce IncRec, a retrieval framework that aims to learn extreme cross-scenario incremental samples and align with the preference of ranking stage. In particular, Section ~\ref{sec:isc} describes the cross-scenario incremental sample construction; Section ~\ref{sec:islf} elaborates the architecture of the proposed model in detail.

\subsection{Incremental Sample Construction} ~\label{sec:isc}
\begin{figure}[t] 
	\centering
	\includegraphics[width=\linewidth]{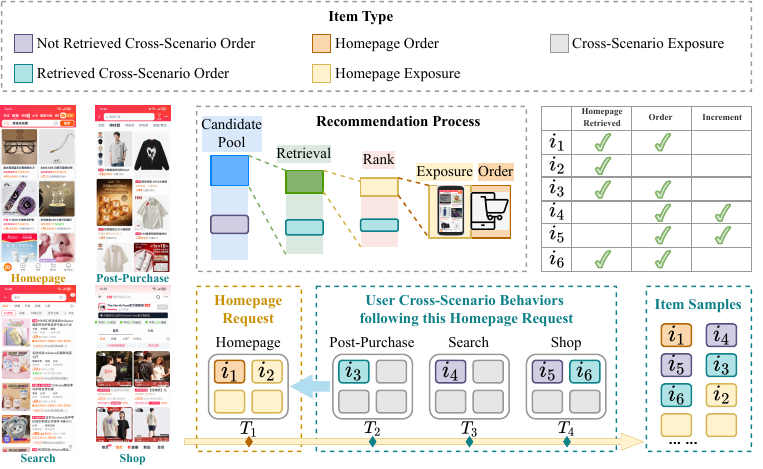}
    \caption{Cross-scenario incremental sample construction}
	\label{theModel}
\end{figure}

The construction pipeline of cross-scenario incremental samples is shown in Figure \ref{theModel}. Given a user $u$, $\mathbf{P}$ represents the item candidate pool, $\mathbf{R}^u = \{i_1,i_2,i_3,i_6,...\}$ denotes the set of results retrieved by multiple models from $\mathbf{P}$ with respect to the request on homepage recommendation at time $T_1$, $\mathbf{X}^u=\{i_1,i_3,i_4,i_5,i_6\}$ represents user cross-scenario behaviors after time $T_1$.
While our primary focus is on serving the homepage recommendation scenario, we integrate behaviors from other scenarios to capture more comprehensive user interests. Therefore, the target items to be predicted include not only the homepage-ordered $i_1$, but also the subsequent $i_3$, $i_4$, $i_5$, $i_6$ ordered in other scenarios after time $T_1$. It is crucial to note that the unretrieved items $i_4$ and $i_5$ in this homepage request constitute potential incremental gains for homepage recommendation. To pay more attention to the optimization of these items, we partition the target items $\mathbf{X}^u$ into two groups:

\begin{itemize}

\item{RTG (Retrieved Target Group): Contains target items already retrieved by existing models (e.g., $i_1$, $i_3$, $i_6$)}
\item{ITG (Incremental Target Group): Contains target items unretrieved by existing models (e.g., $i_4$, $i_5$), representing the core focus of this work.}
\end{itemize}

\subsection{Incremental Sample Learning Framework} ~\label{sec:islf}
\begin{figure}[t] 
	\centering
	\includegraphics[width=\linewidth]{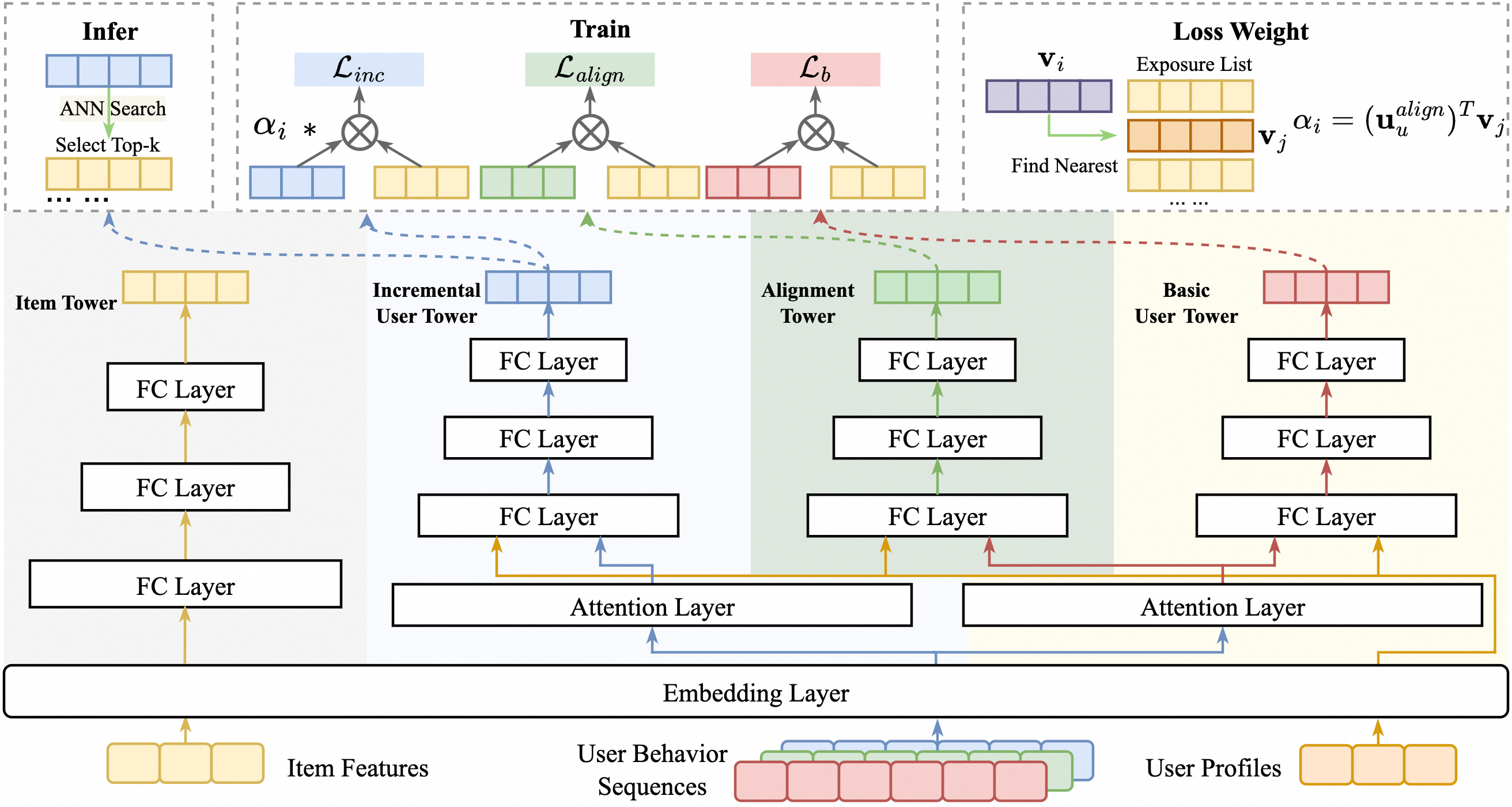}
    \caption{Architecture of IncRec. The model employs a shared item tower and three specialized user towers: the basic tower captures overall user interests, the incremental tower focuses on incremental interests, and the alignment tower ensures consistency with downstream ranking preference.}
	\label{theModelFrame}
\end{figure}

In this paper, we emphasize the critical role of incremental samples in improving the overall retrieval performance. The limited quantity of incremental samples hinders model convergence and impedes the learning of sufficiently discriminative representations. Therefore, we adopt a shared-bottom multi-task training paradigm. The framework of our model is shown in Figure ~\ref{theModelFrame}. A basic dual-tower architecture is employed to capture comprehensive user representations and learn latent parameters sufficiently from both RTG and ITG. Then an incremental user tower is introduced to further extract incremental representations from ITG via explicit contrast. In addition, it is worth noting that not all incremental items can be accepted by ranking stage. From the retrieval system's perspective, items endorsed by ranking stage yield greater systemic benefits. Thus, a consistency-aware alignment tower is introduced to ensure consistency between retrieval and ranking.

\subsubsection{\textbf{Basic Dual-Tower Architecture}} To mitigate the impact of limited incremental samples, we build a basic user tower and item tower that use samples from both RTG and ITG. The basic user representation and item representation can be obtained by:
\begin{equation}
\mathbf{u}^{b}_u=f_{b}(\mathcal{P}_u, \mathcal{B}_u),\;\mathbf{v}_{i}=g_{item}(\mathcal{I}_i),
\end{equation}
where $f_{b}$ represents arbitrary user model, $g_{item}$ represents item model, $\mathcal{P}_u$, $\mathcal{B}_u$, $\mathcal{I}_i$ represents user profile, user historical behavior sequence and target item feature respectively. The similarity between a user $u$ and a item $i$ can be estimated by the inner product $(\mathbf{u}^{b}_u)^T\mathbf{v}_i$. Following the widely used embedding-based retrieval method, we can optimize the similarity using contrastive learning. The loss function is defined as follows: 
\begin{equation}
\mathcal{L}_b=-\sum_u \sum_{i \in RTG \cup ITG}log \dfrac{sim_b(u,i)}{sim_b(u,i) + \sum_{j\in \mathcal{N}}sim_b(u,j)},
\end{equation}
where $sim_b(u,i)=\exp((\mathbf{u}^{b}_u)^T\mathbf{v}_i)$, $\mathcal{N}$ denotes a set of negative samples randomly sampled from item candidate pool. 
\subsubsection{\textbf{Incremental User Tower}} To better capture incremental user representations that are under-explored by existing models, we introduce an incremental user tower, which uses the same model structure as the basic user tower but with different parameters. Then the incremental user representation is obtained by:
\begin{equation}
\mathbf{u}^{inc}_u=f_{inc}(\mathcal{P}_u, \mathcal{B}_u),
\end{equation}
It is worth noting that retrieving items that are already retrieved by other models is meaningless for entire system. Therefore, the main objective of incremental user tower is to distinguish between items from ITG and items from RTG. In other words, the incremental user tower takes items from ITG as positive samples and items from RTG as negative samples. The loss function can be defined as: 
\begin{equation} \label{eq1}
\mathcal{L}_{inc}=-\sum_u \sum_{i \in ITG} \alpha_{i} log \dfrac{sim_{inc}(u,i)}{sim_{inc}(u,i) + \sum_{j\in RTG \cup \mathcal{N}}sim_{inc}(u,j)},
\end{equation}
where $sim_{inc}(u,i)=\exp((\mathbf{u}^{inc}_u)^T\mathbf{v}_i)$, $\alpha_{i}$ denotes the adaptive consistency weight of each item $i$ in ITG, which is computed in the following section. Higher $\alpha_{i}$ means greater exposure probability, which consequently guides the incremental user tower to pay more attention to the optimization of this item.
\subsubsection{\textbf{Consistency-aware Alignment Tower}} \label{sec:align tower} Due to the significant differences in user interest distributions across different scenarios, many cross-scenario incremental samples may not be accepted by ranking stage. It is crucial for the model to focus more on incremental samples that are more consistent with ranking preference. To achieve this, we introduce an additional alignment tower that focuses on learning consistency task and transferring consistency capability to the incremental tower. Alignment tower also shares the same structure as the basic user tower but with different parameters. The alignment user representation is obtained by:
\begin{equation}
\mathbf{u}^{align}_u=f_{align}(\mathcal{P}_u, \mathcal{B}_u),
\end{equation}
To keep better consistency with ranking, it takes the exposed items (ETG) as positive samples, and randomly sampled items as negative samples. Then the loss function of alignment tower can be defined as:
\begin{equation}
\mathcal{L}_{align}=-\sum_u \sum_{i \in ETG}log \dfrac{s_{align}(u,i)}{s_{align}(u,i) + \sum_{j\in \mathcal{N}}s_{align}(u,j)},
\end{equation}
where $s_{align}(u,i) = \exp((\mathbf{u}^{align}_u)^T\mathbf{v}_i)$.
It is important to transfer the consistency capability to the incremental user tower and make model pay more attention to incremental items with high exposure probability. However, it is not feasible to directly use the inner product between the alignment user representation and item representations as the consistency weight for incremental samples in the incremental tower. This is because the incremental samples have not been exposed in homepage and not been trained by alignment tower, leading to biased scoring results. To address this issue, we are of the opinion that similar items should have similar rank scores. Therefore, we first select the most similar item $j$ to each incremental item $i$ from the exposed items based on representation similarity. Then we use inner product between item $j$'s representation and alignment user representation as the consistency weight $\alpha_{i}$ defined in Equation ~\ref{eq1}:
\begin{equation}
\label{findsim}
\begin{split}
& j = \mathop{\arg\max}\limits_{j \in ETG} (\mathbf{v}_i)^T\mathbf{v}_j, \\
& \alpha_{i} = (\mathbf{u}^{align}_u)^T\mathbf{v}_j
\end{split}
\end{equation}
In other words, we assign greater adaptive weight to incremental sample with higher exposure probability, to achieve consistency alignment with ranking stage.
\subsubsection{\textbf{Optimization and Prediction}}
The overall loss of the model is defined as :
\begin{equation}
\mathcal{L}=\mathcal{L}_{b}  + \mathcal{L}_{inc} + \mathcal{L}_{align}.
\end{equation}
Then we choose AdaGrad to train the model. After training, we use the item tower to infer item representations for large-scale candidate pool, and we use the incremental user tower to infer user representation. Then the user representation is utilized to retrieve top-K items using an approximate nearest neighbor approach ~\cite{ann}.
\section{EXPERIMENT}
In this section, we first give the experimental setups, and then apply our framework to representative retrieval methods for comparative experiments. Next, we conduct ablation study on IncRec to validate the importance of our proposed module. Finally, we perform online A/B tests on Taobao to evaluate its effectiveness in production environments.
 


\begin{table*}[ht]
\vspace{-5pt}
  \caption{Results of performance comparison} \vspace{-5pt}
  \centering
  \label{pc}
    {
    \small
    \begin{tabular}{p{1.8cm} p{0.9cm} p{0.9cm} p{0.9cm} p{0.9cm} p{0.9cm} p{0.9cm} p{0.9cm} p{0.9cm} p{0.9cm} p{0.9cm} p{0.9cm} p{0.9cm}}
      \toprule
       & \multicolumn{3}{c}{\textbf{Kuairand Hit@2k}} & \multicolumn{3}{c}{\textbf{Kuairand Hit@4k}} & \multicolumn{3}{c}{\textbf{Taobao Hit@4k}} & \multicolumn{3}{c}{\textbf{Taobao Hit@4k}} \\
       \cmidrule(lr){2-4} \cmidrule(lr){5-7} \cmidrule(lr){8-10} \cmidrule(lr){11-13}
      \textbf{Model} & {Base@1k} & {Sup@1k} & {Inc@1k} & {Base@2k} & {Sup@2k} & {Inc@2k} & {Base@1k} & {Sup@1k} & {Inc@1k} & {Base@1k} & {Sup@1k} & {Inc@1k}\\
      \midrule
        {YouTube-DNN} & {1.33\%} & 0.65\% & \underline{\textbf{2.21\%}} & 1.98\% & {0.96\%} & \underline{\textbf{3.27\%}} &{0.94\%} & 0.83\% & \underline{\textbf{0.91\%}} & 1.77\% & {1.37\%} & \underline{\textbf{1.42\%}}\\
    \midrule
    {DSSM}& {1.37\%} & 0.63\% & \underline{\textbf{3.41\%}} & 2.00\% & {1.01\%} & \underline{\textbf{5.85\%}} & {0.86\%} & 0.72\% & \underline{\textbf{1.40\%}} & 1.57\% & {1.31\%} & \underline{\textbf{2.54\%}}\\    
    \midrule
    {SASRec} & {1.15\%} & 0.59\% & \underline{\textbf{3.30\%}} & 1.74\% & {1.02\%} & \underline{\textbf{5.81\%}} & {0.91\%} & 0.70\% & \underline{\textbf{1.43\%}} & 1.60\% & {1.38\%} & \underline{\textbf{2.57\%}}\\  
    \midrule
    {Comirec} & {1.17\%} & 0.49\% & \underline{\textbf{2.24\%}} & 1.67\% & {0.70\%} & \underline{\textbf{3.39\%}} & {0.93\%} & 0.51\% & \underline{\textbf{0.81\%}} & 1.44\% & {0.95\%} & \underline{\textbf{1.50\%}} \\  
    \midrule
    {Kuaiformer} & {1.56\%} & 0.71\% & \underline{\textbf{2.52\%}} & 2.27\% & {0.88\%} & \underline{\textbf{4.64\%}} & {1.33\%} & 0.98\% & \underline{\textbf{1.18\%}} & 2.31\% & {1.21\%} & \underline{\textbf{2.16\%}}\\ 
      \bottomrule
    \end{tabular}}
  \vspace{-10pt}
  \label{tab:merged_rows}
\end{table*}

\noindent \textbf{Dataset.} 
We adopt the publicly available KuaiRand dataset ~\cite{kuairand} as one of the experimental datasets. In addition, we constructed a comprehensive benchmark dataset based on user interaction logs collected from the Taobao App over 21 consecutive days. The data from the first 20 days were used for training, and the final day was designated as the test set. The dataset includes not only items exposed or clicked on the homepage, but also items purchased in other scenarios following each homepage visit, thereby capturing rich cross-scenario behavioral patterns.

\noindent \textbf{Evaluation metric.}  We adopt incremental hitrate to evaluate the effectiveness of all models for the entire retrieval system, which is defined as:
\begin{equation} \label{hitrate}
	Hitrate_{inc}@K = \sum_{u} \frac{|\mathbf{C}^u \cap ITG|}{|ITG|}  ,
\end{equation}
where $\mathbf{C}^u$ represents the retrieved top-K items for user $u$.
\noindent \textbf{Baselines.} To evaluate the effectiveness of IncRec, we integrate it into five representative retrieval models, including YouTube-DNN ~\cite{towtower3}, DSSM ~\cite{dssm}, SASRec ~\cite{sasrec},
ComiRec ~\cite{twotower2}, KuaiFormer ~\cite{kuaiformer}, and Taobao online model (TB-Online).

\subsection{Performance Comparison}
We utilize the non-feed portion of the KuaiRand dataset as cross-scenario data. To simulate existing retrieval framework, we first apply item-to-item method over all samples to initialize the baseline retrieval state. Samples that are not retrieved by the item-to-item method are identified as incremental samples. These samples are then used to evaluate both representative retrieval models and their IncRec-enhanced variants in terms of hitrate.

Table ~\ref{pc} compares the base model with its IncRec-enhanced variant, where the best results are highlighted in \textbf{bold} and \underline{underlined}. For comparing offline performance more fairly, we remove the alignment module of IncRec. Base@1k represents the hitrate of the top 1k items from the base model, and Sup@1k represents the hitrate of the items ranked 1k - 2k retrieved by the base model. Inc@1k denotes the hitrate of the top 1k items retrieved by the enhanced model after deduplicating the results from Base@1k. All data in Table ~\ref{pc} represent absolutely incremental performance gains.

As shown, when enhanced with IncRec, both Inc@1k and Inc@2k of these six models significantly outperform the Sup@1k and Sup@2k of their base models. Notably, DSSM and SASRec achieve substantial gains in Inc@1k relative to their corresponding Sup@1k scores. In addition, comparing Base@2k and Sup@2k, we can observe that expanding the base model leads to a significant decrease in incremental hitrate. On the contrary, Inc@2k still approaches even outperforms Base@2k on some models. The universal improvement across diverse retrieval models validates and superiority of IncRec. These results position IncRec as a universally applicable enhancement strategy for industrial recommendation systems.




\subsection{Ablation Study}

To clearly illustrate the contribution of the core components, we perform two ablation experiments on the IncRec framework.


\begin{table}[ht]
\vspace{-5pt}
  \caption{Results of ablation study on incremental tower} \vspace{-5pt}
  \centering
    {\small
    \begin{tabular}
    {ccccc }
      \toprule
       
       & \multicolumn{2}{c}{\textbf{Order Hit@2k}} & \multicolumn{2}{c}{\textbf{Order Hit@4k}}  \\
       \cmidrule(lr){2-3} \cmidrule(lr){4-5} 
      \textbf{Model} & {Base@1k} & {Inc@1k} & {Base@2k} & {Inc@2k}  \\
      \midrule
      TB-Online
        & \multirow{4.5}{*}{1.44\%} 
        & 0.87\% 
        & \multirow{4.5}{*}{2.30\%} 
        & 1.45\% 
         \\
        \cmidrule(lr){1-1} \cmidrule(lr){3-3} \cmidrule(lr){5-5}
        TB-ITG && 1.42\% && 2.41\%  \\
        TB-boost && 1.38\% && 2.73\% \\
        IncRec-NA && \underline{\textbf{2.15\%}} && \underline{\textbf{3.49\%}}  \\

        
      \bottomrule
    \end{tabular}}
  \vspace{-10pt}
  \label{aa}
\end{table}

\textbf{The impact of incremental learning tower.} The results are shown in Table~\ref{aa}. The baseline is the Taobao online dual-tower model, trained with RTG and ITG samples. TB-boost applies amplified weights to ITG samples, while TB-{ITG} exclusively utilizes ITG samples during training. IncRec-NA extends TB-Online by incorporating an additional incremental learning tower (corresponding to IncRec without alignment module). On average, TB-ITG and TB-boost achieve 60\% and 77\% relative improvements over TB-Online in terms of Inc@1k and Inc@2k respectively, highlighting the critical role of ITG samples. The introduction of incremental learning tower further yields additional 53\% and 44\% relative enhancements in Inc@1k and Inc@2k, confirming the effectiveness of the proposed IncRec.

\begin{table}[ht]
\vspace{-5pt}
  \caption{Results of ablation study on alignment tower} \vspace{-5pt}
  \centering
    {\small
    \begin{tabular}
    {p{1.3cm} p{0.79cm} p{0.79cm} p{0.79cm} p{0.79cm} p{0.79cm} p{0.79cm}}
      \toprule
       
       & \multicolumn{2}{c}{\textbf{Order Hit@2k}} & \multicolumn{2}{c}{\textbf{Order Hit@4k}} & \multicolumn{2}{c}{\textbf{Exposure Hit}} \\
       \cmidrule(lr){2-3} \cmidrule(lr){4-5} \cmidrule(lr){6-7}
      \textbf{Model} & {Base@1k} & {Inc@1k} & {Base@2k} & {Inc@2k} & {Inc@1k} & {Inc@2k} \\
      \midrule
        IncRec-NA & \multirow{3}{*}{1.44\%}  & \underline{\textbf{2.15\%}} & \multirow{3}{*}{2.30\%} &
        \underline{\textbf{3.49\%}} & 2.44\% & 3.32\% \\
        \cmidrule(lr){1-1} \cmidrule(lr){3-3} \cmidrule(lr){5-7}

        IncRec-ori && 1.54\% && 2.69\% & 3.27\% & 3.95\% \\
        IncRec && 1.66\% && 2.92\% & \underline{\textbf{3.41\%}} & \underline{\textbf{4.35\%}} \\
        
      \bottomrule
    \end{tabular}}
  \vspace{-10pt}
  \label{aa2}
\end{table}

\textbf{The impact of consistency-aware alignment tower.} The results are displayed in Table ~\ref{aa2}. The Exposure Hit metric refers to the proportion of exposed items that are retrieved,  indicating the ability in aligning with the preference of downstream stage. Both IncRec-ori and IncRec introduce a consistenct-aware alignment tower. IncRec-ori computes the adaptive weight based on the similarity between alignment user representation and incremental item representation, while IncRec uses implicit adaptive weight derived from Equation ~\ref{findsim}. As is shown, IncRec-ori and IncRec significantly outperform IncRec-NA in exposure hitrate by a large margin, despite slight reductions in order hitrate. It is crucial to note that IncRec outperforms IncRec-ori in both order hitrate and exposure hitrate, verifying the effectiveness of the proposed alignment method.

\subsection{Online A/B Test}
\begin{table}[ht]
\vspace{-5pt}
  \caption{Results of online A/B test} \vspace{-5pt}
  \label{ol}
  \label{tab:freq}
  \begin{tabular}{ccccc}
    \toprule
    model & hitrate & pvr & transaction count \\
    \midrule
    IncRec-NA & +10.20\% & 5.78\% & +0.9\% \\
    IncRec & +9.07\% & 7.73\% & +1.2\% \\
    \bottomrule
  \end{tabular}
  \vspace{-10pt}
\end{table}
To quantify the online performance of IncRec, we conduct strict online A/B tests on Taobao App's homepage recommendation scenario with 1\% main traffic. We use online hitrate to evaluate the effect of IncRec on overall retrieval, and use PVR (Proportional Visibility Rate) and transaction count to evaluate its effect on whole homepage recommendation. The results are shown in Table~\ref{ol}. As shown, both IncRec-NA and IncRec can notablely improve the overall retrieval hitrate by a large margin, validating the value of IncRec. In addition, IncRec achieves 33\% higher PVR compared to IncRec-NA, demonstrating the advantage of aligning with the preference of ranking. Finally, IncRec-NA and IncRec achieve 0.99\% and 1.2\% improvements in transaction count respectively. This indicates
that the proposed framework achieves much better recommendation results and
delivers considerable increments for the platform.





\section{CONCLUSION}
In this work, we propose a novel retrieval framework IncRec for learning cross-scenario incremental samples that are not retrieved by any existing model within a multi-retrieval system. Extensive offline evaluations and online A/B tests demonstrate the superiority and applicability of our framework in delivering incremental performance gains for the entire retrieval system.
\bibliographystyle{ACM-Reference-Format}
\bibliography{sample-base}










\end{document}